\documentclass[oldversion]{aa}

\pdfoutput=1

\usepackage{graphicx}
\usepackage{natbib}
\usepackage{textcomp}
\usepackage{amsmath}
\usepackage{amssymb}
\usepackage{wasysym}
\usepackage{fancybox}
\usepackage{multirow}
\usepackage[rightcaption]{sidecap}

\newcommand{\co}[0]{CoRoT}

\newcommand{\fe}[0]{pdf}

\begin{document}

\title{Planetary eclipse mapping of CoRoT-2a}
\subtitle{Evolution, differential rotation, and spot migration}

\author{K. F. Huber
   \and S. Czesla
   \and U. Wolter
   \and J. H. M. M. Schmitt}

\institute{Hamburger Sternwarte, Universit\"at Hamburg, Gojenbergsweg 112, 21029 Hamburg, Germany}

\date{Received ... / Accepted ... }

\abstract {The lightcurve of \co-2 shows substantial rotational modulation 
           and deformations of the planet's transit profiles caused by starspots. 
           We consistently model the entire lightcurve, including both rotational modulation and transits,
           stretching over approximately $30$~stellar rotations and $79$~transits.
           The spot distribution and its evolution on the noneclipsed and eclipsed surface sections are presented
           and analyzed, making use of the high resolution achievable under the transit path. \\
           We measure the average surface brightness on the eclipsed section to be ($5\pm1$)~\% lower than on the noneclipsed section.
           Adopting a solar spot contrast, the spot coverage on the entire surface reaches up to $19$~\%
           and a maximum of almost $40$~\% on the eclipsed section.
           Features under the transit path, i.e. close to the equator, rotate with a period close to $4.55$~days.
           Significantly higher rotation periods are found for features on the noneclipsed section
           indicating a differential rotation of $\Delta \Omega > 0.1$.
           Spotted and unspotted regions in both surface sections concentrate on preferred longitudes
           separated by roughly $180$\textdegree.
           }


\maketitle

\section{Introduction}
\label{Sec:Intro}

The space-based \co \ mission \citep[e.g.][]{Auvergne2009},
launched in late~2006, provides stellar photometry of
unprecedented quality.
One of CoRoT's primary tasks is the search for extra-solar planets
using the transit method.
So far several systems with eclipsing exoplanets were found,
one of them \co-2 harboring a giant close-in 'hot-jupiter'.

The planet \co-2b was discovered by \citet{Alonso2008}, who
determined the system parameters from transits and follow-up radial
velocity (RV) measurements.
\citet{Bouchy2008} observed the Rossiter-McLaughlin effect using additional
RV measurements and determined the projected angle \mbox{$\lambda \, = \, (7.2 \, \pm \, 4.5)$\textdegree}
between the stellar spin and the planetary orbital axis.
\citet{Lanza2009} used the strong rotational modulation of the lightcurve to study
the spot distribution and its evolution; they detect a stellar rotation
period of \mbox{$P_{rot} \, = \, (4.522 \, \pm \, 0.024)$~d} and two active longitudes
on opposite hemispheres.
The secondary eclipse of the planet was first detected by \citet{Alonso2009} in white light.
Later the analysis was refined and extended by \citet{Snellen2009},
detecting significant thermal emission of the planet;
\citet{Gillon2009} repeated this work using additional infrared data.
They also find an offset of the secondary transit timing
indicating a noncircular orbit.

The determination of the planetary parameters by \citet{Alonso2008} did not account
for effects of stellar activity, which is in general not negligible for active stars.
\citet{Czesla2009} re-analyze the transits and derive new parameters for the planet radius~$R_p/R_s$
and the orbital inclination~$i$ considering the deformation of transit profiles due to spot occultation.
This is especially important for attempts to reconstruct active surface regions 
from transit profiles. 
An analysis of a single transit by \citet{Wolter2009} shows the potential of eclipse-mapping
and yielded constraints on the properties of the detected starspot.
Similar approaches to analyze signatures of starspots in transit profiles
were also carried out by \citet{Pont2007} (HD~$189733$) and \citet{Rabus2009} (TrES-$1$)
primarily using data obtained with the Hubble Space Telescope.
In a more comprehensive approach,
\citet{Huber2009} reconstructed an interval of the \mbox{\co-2} lightcurve
over two stellar rotations including the transits.

This work is based on the paper of \citet{Huber2009}.
We now analyze the \textit{entire} data set of \co-2,
modeling both the rotational modulation of the global
lightcurve and transits simultaneously.
In this way we derive stellar surface maps for both the noneclipsed
and the eclipsed section of the star. 

\section{Observations and data reduction}
\label{Sec:Object}

\begin{table}[t]
  \begin{minipage}[h]{0.5\textwidth}
    \renewcommand{\footnoterule}{}
    \caption{Stellar/planetary parameters of CoRoT-2a/b. \label{Tab:Exo2prop}}
    \begin{center}
      \begin{tabular}{l c c}
      \hline \hline
      Star \hspace*{0.0cm} \footnote{$P_s$ - stellar rotation period}
                     & Value $\pm$ Error & Ref.\footnote{taken from \citet{Lanza2009} [L09],
                                                         \citet{Alonso2008} [A08], \citet{Bouchy2008} [B08], or \citet{Czesla2009} [C09]} \\
      \hline
      $P_s$          & $(4.522 \, \pm \, 0.024)$ d         & L09 \\
      Spectral type  & G7V                                 & B08 \\
      \hline \hline
      Planet \hspace*{0.0cm} \footnote{$P_p$ - orbital period,
                                       $T_c$ - central time of first transit,
                                       $i$ - orbital inclination,
                                       $R_p, R_s$ - planetary and stellar radii,
                                       $a$ - semi major axis of planetary orbit,
                                       $u_a, u_b$ - linear and quadratic limb darkening coefficients.}
                                                          & Value $\pm$ Error & Ref.\\
      \hline
      $P_p$       & $(1.7429964 \, \pm \, 0.0000017)$ d       & A08 \\
      $T_c$ [BJD] & $(2\,454\,237.53362 \, \pm \, 0.00014)$ d & A08 \\
      $i$            & $(87.7 \, \pm \, 0.2)$\textdegree      & C09 \\
      $R_p/R_s$   & $(0.172 \, \pm \, 0.001)$                 & C09 \\
      $a/R_s$     & $(6.70 \, \pm \, 0.03)$                   & A08 \\
      $u_a, u_b$      & $(0.41\pm0.03), (0.06\pm 0.03)$       & A08 \\
      \hline
      \end{tabular}
    \end{center}
  \end{minipage}
\end{table}

The data were obtained in the first \textit{long run}
of the \co \ satellite \mbox{(May 16} to \mbox{Oct. 15, 2007)}.
The planetary system \co-2 (Star CorotID 0101206560)
consists of an active solar-like G7V star
and a large planetary companion on a close orbit.
Due to its edge-on view and an orbit period of only about $1.7$~days,
this lightcurve contains about $80$~transits, roughly $3$ during each stellar rotation.

The large surface inhomogeneities of this very active star are clearly visible in the
rotational modulation of the \co \ lightcurve taken over approximately $140$~days.
Although there is surface evolution even on timescales of one stellar rotation,
these changes are small compared to the modulation amplitude.
Significant deformations of the transit profiles due to spots are also visible
throughout the whole time series.

The stellar and planetary parameters used throughout this analysis
are given in Tab.~\ref{Tab:Exo2prop}.
The extensive raw data analysis and reduction follows the descriptions in
\citet{Czesla2009} and \citet{Huber2009}.

In this paper we only analyze the \textit{alarm mode} data of \co-2,
when the satellite switched from a sampling rate of $1/512$~s$^{-1}$ to $1/32$~s$^{-1}$
(which started after $3$~transits),
and use the combination of the three color channels (`white' light).
Our analysis starts after one stellar rotation at
\mbox{JD$\, = \, T_c \, + \, P_s \, = \, 2454242.05562$} (see Tab.~\ref{Tab:Exo2prop})
or a stellar rotation phase of \mbox{$\phi_s \, = \, 1.0$}, respectively.
We use units of stellar rotation phase in our analysis,
the last point of our data interval corresponds to \mbox{$\phi_s \, = \, 31.24$}.
This leaves us with an observation span of more than $30$ stellar rotations containing $79$~transits.
In the following the first transit inside our data interval is labeled $0$, the last $78$.

\section{Analysis}
\label{Sec:Analysis}

 \subsection{The Model}
  The projected axes of the planetary orbit and the stellar rotation are co-aligned \citep{Bouchy2008},
  which strongly suggests a 3-dimensional alignment.
  Hence, the rotation axis of \co-2a is inclined by approximately $88$\textdegree.
  While this impedes the reconstruction
  of latitudinal information of surface features,
  the existence of a planet crossing the stellar disk
  allows to access latitudinal information on spots beneath its path.
  During a planetary passage the surface brightness distribution is mapped onto
  the lightcurve as deformations of the transit profiles.
  As a consequence of the co-aligned orientation of the planetary orbit and the stellar spin,
  the surface band scanned by the planetary disk remains the same:
  the planet constantly crosses the latitudinal band between $6$\textdegree \ and $26$\textdegree.
  Accordingly, the stellar surface can be subdivided into two
  sections: the eclipsed section and the noneclipsed section \citep[cf.,][]{Huber2009}.
  
  Our surface model subdivides the two individual sections into a number of `strips';
  $N_e$ is the number of strips in the eclipsed and $N_n$ in the noneclipsed section, respectively.
  Each strip represents a longitudinal interval
  inside the latitudinal boundaries of the corresponding section.
  The layout of our surface model is shown in Fig.~$1$ of \citet[][]{Huber2009}.

 \subsubsection{Model resolution and error estimation}
 \label{Sec:RESandERR}

  The problem of lightcurve inversion is well known to be ill-posed,
  so that the parameter space is usually further constraint by a regularization.
  One such regularization is the maximum entropy approach, applied, for example, by
  \citet{Lanza2009} in their analysis of CoRoT-2a.

  We use a Nelder-Mead (NM) Simplex algorithm for minimization \citep{NR-BOOK}.
  Our model does not require any regularization because of its relatively small number of parameters.
  As discussed by \citet{Huber2009}, we choose a number of strips balancing
  the improvement in $\chi^2$ and the deterioration in uniqueness.
  For higher strip numbers, adjacent strips increasingly influence each other
  because brightness can be redistributed without significant loss of fit quality.
  In our error analysis we assume that there is a unique best-fit solution to our problem and that
  the NM algorithm approaches it to within the limits of its ability to converge.
  Starting from the yet unknown best-fit solution, brightness can be redistributed among
  the strips at the expense of fit quality. 
  A set of Nelder-Mead fit runs will, therefore, provide a sequence of solutions with
  different realizations of this brightness redistribution.

  We calculate $50$~reconstructions with randomized starting points and
  adopt the average of all reconstructions as our most appropriate model.
  As an estimate for the error of the strip brightness, we use the standard deviation
  of the parameter values, obtained from the set of reconstructions.

  We do point out that the averaged solution appears smoother than most individual reconstructions in the
  sense that the difference between adjacent strip brightnesses is smaller. This can be understood in the
  picture of brightness redistribution, because the brightening of one strip may preferably be compensated
  by darkening an adjacent one, which increases their contrast.
  This effect is largely canceled out by the averaging,
  which makes it appear much like a regularization of the solutions.

  It may be criticized that the brightness error we use is largely determined
  by the ability of the NM algorithm to converge to a unique solution.
  We emphasize, however, that the $\chi^2$ range covered by the $50$ reconstructions
  exceeds that required for `classical' error analysis, and the estimate will, therefore,
  remain rather conservative from that point of view.

  Our test runs indicate that for our purpose the most appropriate number of strips to choose
  for the noneclipsed section is $N_n = 12$, which will be used in our analysis.
  Larger numbers of $N_n$ appear to already oversample the surface significantly.
  The strip number for the eclipsed section is chosen to be $N_e = 24$ which approximately reflects the size of
  features that are resolvable inside of transits \citep{Huber2009}.


\subsection{Normalization}
\label{Sec:norm}

The observations are normalized with respect to the ab initio
unknown spot-free photospheric flux of the star,
which is defined as maximum brightness \mbox{$b_{phot} \, = \, 1$}.
Unfortunately, it is not trivial to obtain this photospheric flux level
because spots are likely to be located on the visible disk at all times.

A possible solution for this problem is to adopt the maximum observed flux as photospheric.
However, this presumably introduces an error
because the brightest part of the lightcurve shows only the flux level
of the stellar disk during the minimum observed spot coverage, which needs not be zero.
\citet{Lanza2009} determine an average minimum of flux deficit of approximately $2.5$~\%.
Tests with slightly varying maximum brightness values in our reconstructions show no
qualitative difference in the brightness distribution except for a change in the average \textit{total} spot coverage;
however, they show a significant decrease in $\chi^2$ of the lightcurve reconstructions
for a maximum brightness $1$~\% to $2$~\% larger than the highest observed flux.
As a result we choose a photospheric flux level of $2$~\% higher than the maximum observed flux.
The entire lightcurve is normalized with respect to this value.

%
%

\begin{figure}
  \centerline{\includegraphics[height=1.28\textwidth,angle=180]{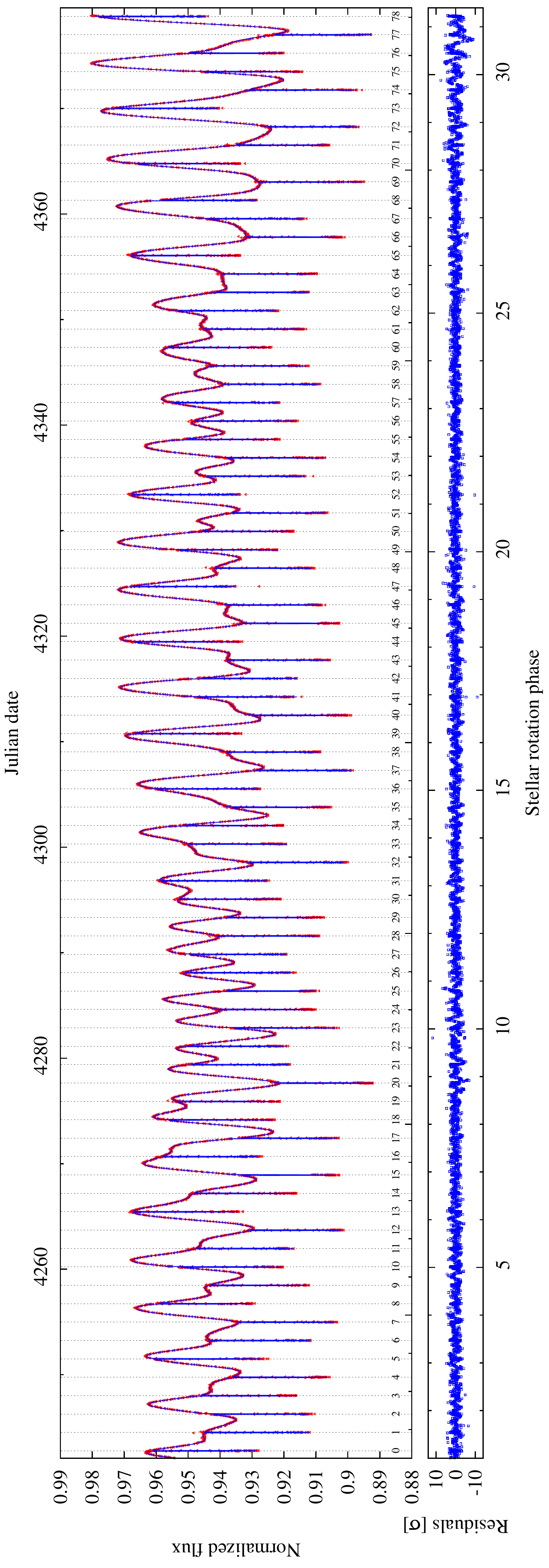}}
  \caption{\label{Fig:allLC}
           Observations and best fit model for \co-2.
           The transit numbers are plotted below the lightcurve;
           for a detailed presentation of the transits see Fig.~\ref{Fig:TRrec}.
           See Sec.~\ref{Sec:res-model} for discussion.}
\end{figure}

\subsection{Lightcurve modeling}
\label{Sec:lc-model}

For analysis, the lightcurve is split into equally sized intervals, each covering one stellar rotation and three transits.
Due to surface evolution detectable on timescales smaller than one stellar rotation,
we choose to define a new interval after each transit.
Thus, the resulting intervals overlap.
Interval~$0$ contains the transits number $0$ to $2$, interval~$1$ contains the transits number $1$ to $3$, and so on.
This way we end up with $76$ lightcurve intervals which are individually reconstructed by our modeling algorithm.

Using fit intervals smaller than one stellar rotation could further reduce the influence of surface evolution.
However, we need at least three transits in each interval to sufficiently cover the eclipsed section.
Therefore, we always use a complete rotation for each reconstruction interval.

We use the fitting method presented in \citet{Huber2009}.
Each fit interval is rebinned to \mbox{$94 \times 32 = 3\,008$~seconds} for the global lightcurve and
to \mbox{$128$~seconds} inside of transits.
Transit points are weighted with a factor of $10$ higher than global points to give them approximately the same
weight in the minimization process.
We assume the planet to be a dark sphere without any emission; this seems to be a good approximation
considering a secondary transit depth of about \mbox{($0.006\,\pm\,0.002$)~\%} \citep{Alonso2009}.
All other necessary parameters can be found in Table~\ref{Tab:Exo2prop}.

We introduce a penalty function to suppress reconstructed brightnesses above 
the photospheric value of one.
Without this boundary the brightness of individual strips exceeds this limit in some reconstructions.
Strips with values above unity must be interpreted as regions with a brightness greater than
the (defined) photosphere;
however, in this approach we want to consider only cool surface features,
which was found to be a good approximation by \citet{Lanza2009}.
Several tests showed that this penalty function does not significantly alter
the outcome of our reconstructions;
it primarily prevents the minimization process from getting stuck in (local)
minima outside the relevant parameter range.

To analyze the transits accurately, we require an `undisturbed transit profile',
which means a profile corrected for the effects of stellar activity.
This `standard transit profile' was determined by \citet{Czesla2009},
where a planet size of \mbox{$R_p/R_s = 0.172$} was found.
With the planetary parameters derived by \citet{Alonso2008}
no satisfactory fits to the transits and global lightcurve can be produced.


\begin{figure}
  \centerline{\includegraphics[scale=0.38,angle=-90]{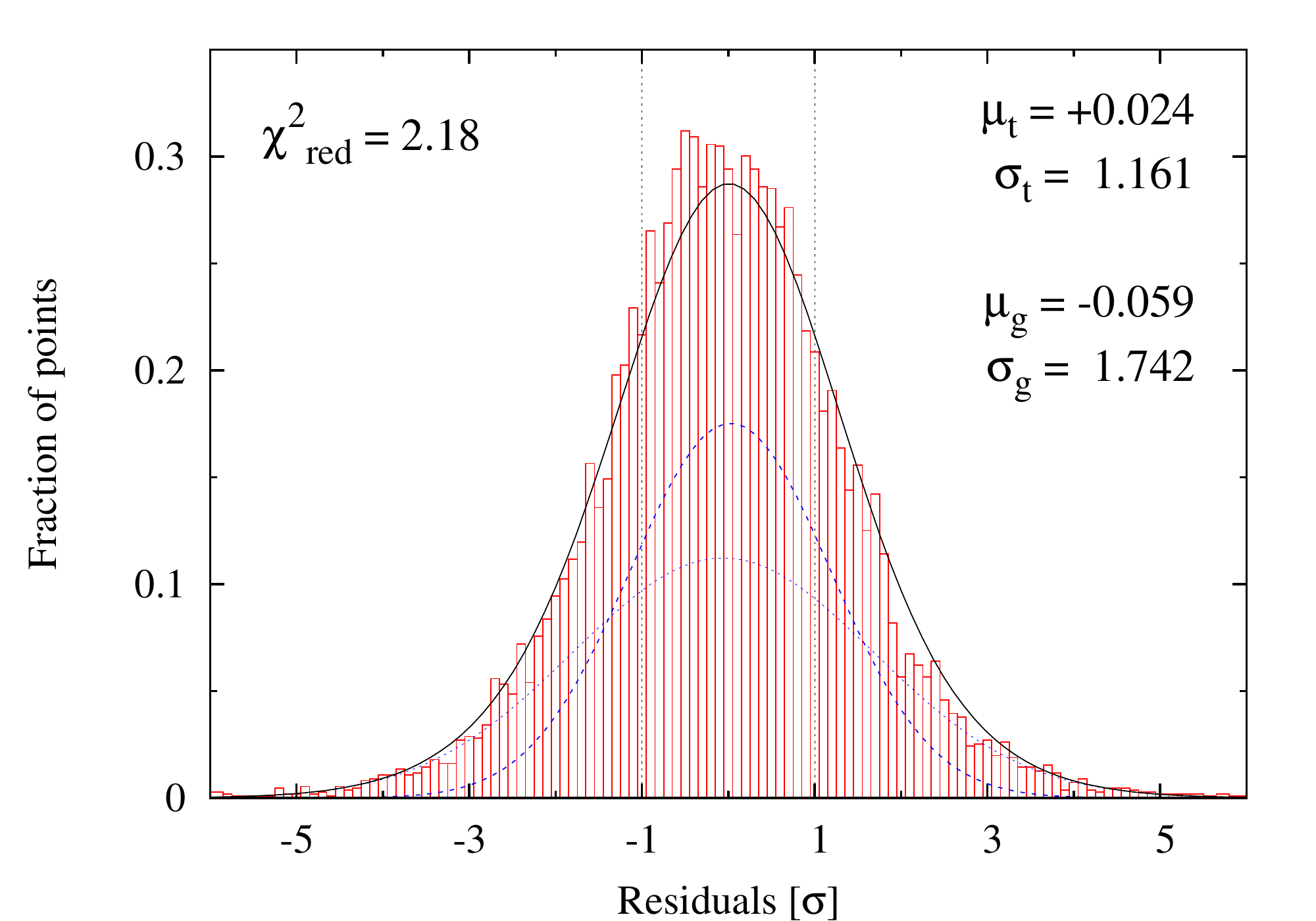}}
  \caption{\label{Fig:err_dist}
           Distribution of residuals $\sigma$ between the observations and the best fit model (see Fig.~\ref{Fig:allLC}).
           The overall distribution is the sum of two Gaussians,
           one coming from inside the transits (mean~$\mu_t$ and width~$\sigma_t$)
           and the other coming from outside of them (global, mean~$\mu_g$ and width~$\sigma_g$).
           For details see Sect.~\ref{Sec:res-model}.
           }
\end{figure}

\subsection{Results of the modeling}
\label{Sec:res-model}

\begin{figure*}
  \centerline{\includegraphics[width=1.0\textwidth,angle=-90]{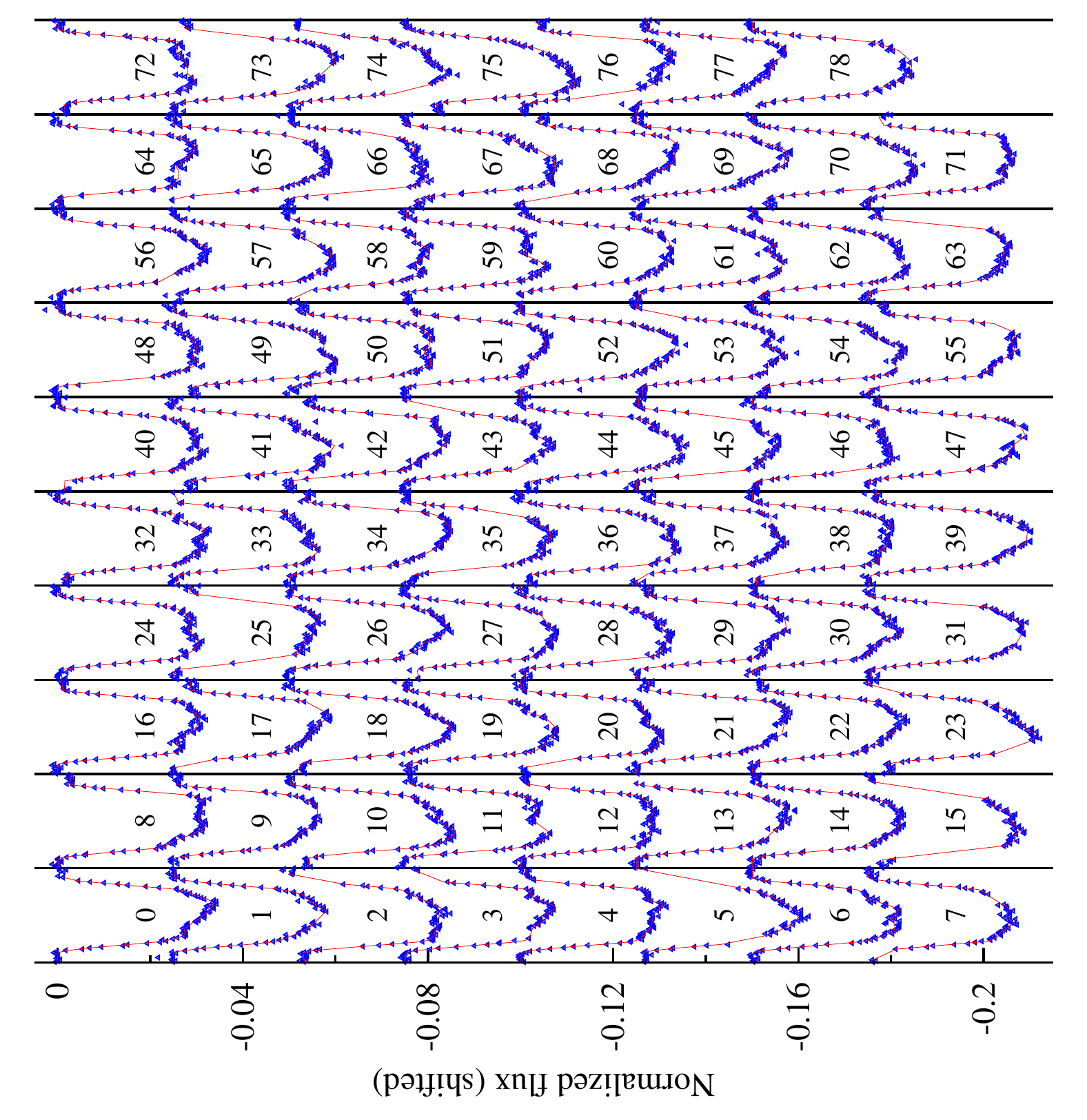}}
  \caption{\label{Fig:TRrec}
           Observation (blue triangles) and reconstruction (red solid line) of $79$~transits from the \mbox{\co-2} lightcurve.
           The first transit of each column is shifted to a continuum of zero, each subsequent transit is shifted by $-0.025$.
           The number of each transit is annotated inside the plot.
           See Fig.~\ref{Fig:allLC} for the entire lightcurve.}
\end{figure*}

We present our lightcurve reconstruction in Fig.~\ref{Fig:allLC}.
It shows the observed \co-2 lightcurve (red triangles), including the transits, and our
reconstruction (blue solid line).
The shown reconstructed lightcurve is a combination of all models for the $76$~fit intervals;
their overlaps were combined using a Gaussian weighting.
Each transit is labeled with its number on the lower edge of the graph;
a more detailed picture of their fits is given in Fig.~\ref{Fig:TRrec},
where we present all transits in a stacked plot.

Below the lightcurve the residuals \mbox{$\sigma \,=\, (O - C) / \sigma_O$}
(\mbox{$O$ - observed lightcurve}, \mbox{$C$ - reconstructed lightcurve}, \mbox{$\sigma_O$ - error of observation})
are given.
The mean values for $\sigma_O$ are $1.8 \cdot 10^{-4}$ outside and $6.9 \cdot 10^{-4}$ inside of transits.
Maximum values of the residuals are approximately $10\,\sigma$,
which corresponds to a deviation of about $0.2$~\% between the observed and reconstructed lightcurves.
On average, the absolute values of residuals are $1.12\,\sigma$ or $O-C \,=\,4.4 \cdot 10^{-4}$, respectively.

Figure~\ref{Fig:err_dist} shows the distribution of residuals from Fig.~\ref{Fig:allLC}.
It is approximately Gaussian, although the underlying distribution is twofold:
one component is a Gaussian distribution of residuals from inside the transits
with \mbox{$\sigma_t = 1.161$}, and another broader component is coming from
the residuals of the global lightcurve, which is significantly wider
with \mbox{$\sigma_g = 1.742$}.
The addition of both Gaussians reproduces the overall distribution of residuals accurately.
We calculate \mbox{$\chi^2_{\mathrm{red}} \,=\, 2.18$}
from all residuals of the entire lightcurve.

The interpretation of $\chi^2_{\mathrm{red}}$ as an actual goodness-of-fit indicator is
not straightforward in this case.
An increase of the strip number should lead towards $\chi^2_{\mathrm{red}}$ values of $1$.
Test calculations show that our fit quality cannot be substantially improved beyond a certain level
when the number of strips is increased.
This level of $\chi^2_{\mathrm{red}}$ primarily reflects the evolution of the lightcurve within one stellar rotation,
which cannot be improved within our static model,
and which is especially visible in the global lightcurve ($\sigma_g \gg \sigma_t$).
Unfortunately, it is very hard to quantify this effect.



\subsection{Construction of brightness maps}
\label{Sec:rec-maps}

\begin{figure*}[t]
 {\large
 \begin{tabular}{c c}
   GLOBAL MAP & TRANSIT MAP  \vspace{-.5cm} \\
 \includegraphics[scale=0.35,angle=-90]{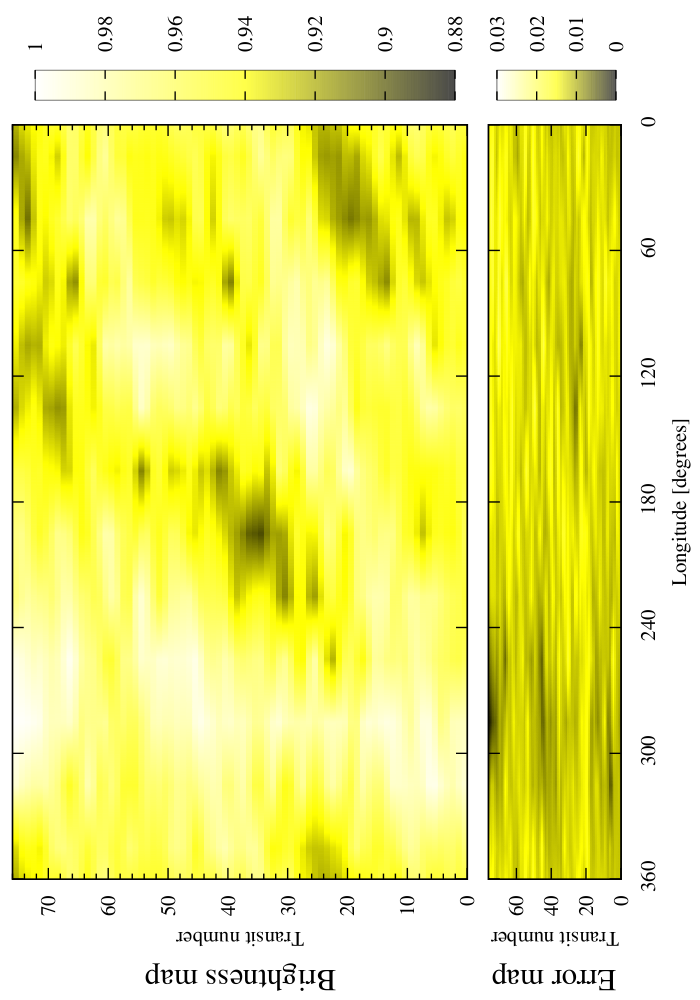} &
 \includegraphics[scale=0.35,angle=-90]{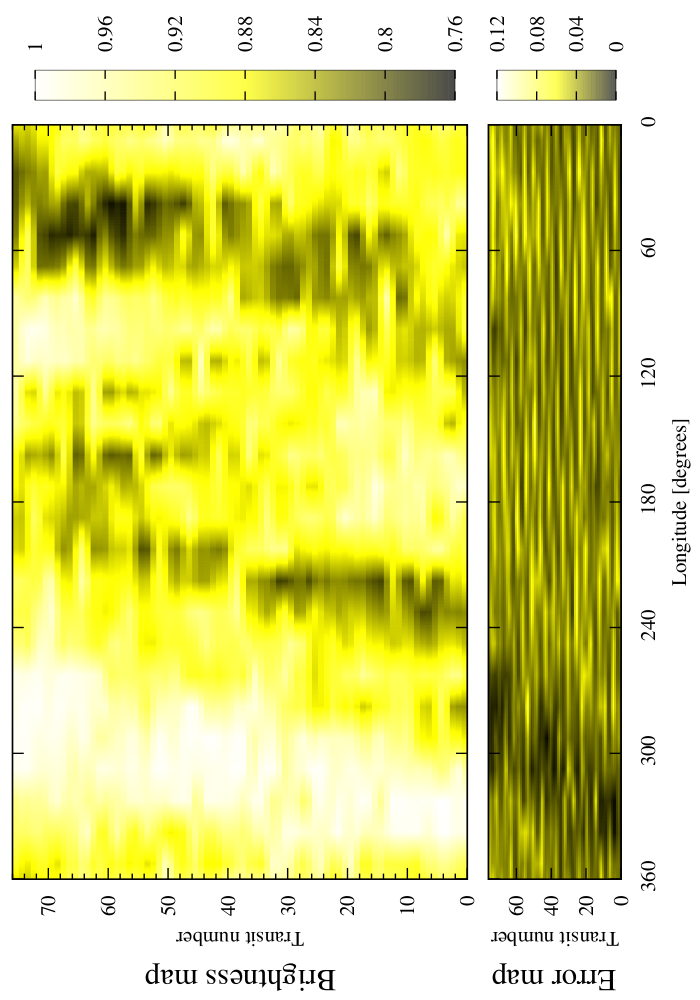} \\
 \end{tabular}
 }
 \caption{\label{Fig:global-transit-maps}
 \textit{Left panel}: Brightness map reconstructed from the global lightcurve (noneclipsed surface).
 \textit{Right panel}: Brightness map for the reconstructions of the transits (eclipsed surface).
 The combination of both maps is presented in Fig.~\ref{Fig:brightnessBOTH}.
 Each row presents the reconstructed brightness distribution of one fit interval;
 the transit number indicates the number of this interval's first transit.
 Each step in transit number equals a temporal step of $1.74$~days.
 The brightness is color-coded, the maximum photospheric brightness is unity.
 The error maps at the bottom of each panel show the estimated reconstruction error.
 See Sects.~\ref{Sec:lc-model} and~\ref{Sec:rec-maps} for details.
 }
\end{figure*}

In Fig.~\ref{Fig:global-transit-maps} we present maps of the temporal evolution
of the brightness distributions.
The left panel gives the brightness map for the noneclipsed,
the right panel for the eclipsed surface section.
The rows show the reconstructed brightness distributions for all $76$ lightcurve intervals;
each interval is labeled by the number of the first transit it contains.
Due to the different resolutions in the two surface sections,
we linearly interpolate the brightness values in each individual row;
there is no interpolation applied between different rows.
To generate the combined brightness map shown in Fig.~\ref{Fig:brightnessBOTH} (left panel),
the maps of the two separated sections are weighted,
corresponding to their disk fraction of $0.79$ for the noneclipsed and $0.21$ for the eclipsed section,
and added.
The errors are combined the same way.
Adjacent rows in our maps are not independent because the fit interval is only shifted by
one transit ($\approx 1/3$ stellar rotation) when moving from one row to the next.

The errors are displayed directly below each map.
For the brightness values of the global lightcurve fit (Fig.~\ref{Fig:global-transit-maps}, left panel),
the mean error is about $1$~\%, for the transits (Fig.~\ref{Fig:global-transit-maps}, right panel)
it is approximately $3$~\%.
The errors in the latter map are larger (on average) and more inhomogeneously distributed
because they also reflect the coverage of the eclipsed section by the transits;
areas only marginally visible in transits cannot be reconstructed with high accuracy.

The stellar longitude scale of our maps runs backwards from $360$\textdegree \ to $0$\textdegree.
This is due to our retrograde definition of the stellar longitude $l$ compared to stellar phases $\phi_s$;
their relation is $l \,=\, (\phi_N - \phi_s) \cdot 360$\textdegree \
(with an integer stellar rotation number $\phi_N \,=\, [\phi_s]+1$).

The identification of significant structures in these maps deserves some attention.
The brightness information is color coded using black color for the darkest structures
and white for photospheric brightness.
Considering the approximate mean error of each map,
we indicate areas with a brightness significantly below unity with yellow color.
Hence, not only black areas of these maps are spots
but yellow structures represent a significant decrease in brightness as well.

\section{Discussion}
\label{Sec:Dis}


In this section we discuss and interpret the spot distributions of our brightness maps
and their evolution.
This involves quite a few different aspects, which are often difficult to disentangle,
and on which we focus on individually in the following subsections.


\subsection{Identifying physical processes and detection limits}

Our brightness maps allow us to witness the evolution of the stellar surface and, to some degree,
to discriminate between individual processes causing changes of the spot distribution:
Emergence/dissociation, differential rotation, or migration of surface features
leave potentially distinguishable signatures.
Unfortunately, the high diversity of \mbox{\co-2a}'s surface
and a limited resolution complicate the interpretation of these signatures.

On the eclipsed section the position of a feature is fairly well known
and it is likely to be physically coherent.
On the noneclipsed section it is not clear whether a feature is actually a single connected active region
or a superposition of two (or even several) different regions.

Features on the eclipsed section provide a valuable reference point for
the detection of differential rotation.
Systematic longitudinal movements of spots in the transit map do not indicate
differential rotation but rather a difference to the input rotation period.
However, a comparison of transit and global maps could reveal
different rotation periods on different surface sections.

Systematic longitudinal movements of structures in the global map do not
necessarily indicate differential rotation either.
A simultaneous decay and growth of two distinct active regions at different longitudes
might leave a signature similar to a single differentially rotating active region.
Therefore, the processes of differential rotation and spot evolution are hard to discern.

A very interesting scenario is the possibility to find spot migration due to the planet.
If a spot moves from the noneclipsed to the eclipsed section, its signature also moves
from the global to the transit brightness map.
If present, such signatures are detectable in high quality brightness maps.

In the following discussion we attempt to attribute signatures
in the brightness maps to the specific \textit{processes} discussed above.
Depending on the number of longitudinal strips used,
map structures have an estimated error in longitude of about half the strip width,
which is about $\pm 7.5$\textdegree \ for the transit and
$\pm 15$\textdegree \ for the global map, respectively.

\subsection{Spot coverage}
\label{Sec:spot-cov}

The brightness maps of Fig.~\ref{Fig:global-transit-maps} show clear evidence of
(a) a coverage of a large fraction of the surface with dark features and
(b) a substantial evolution of this spot distribution
within the roughly $140$~days of observations.
Considering the pronounced rotational modulation of the lightcurve,
and the persistently changing shape between individual rotations,
this result is hardly surprising.
The lowest flux of the rotationally-modulated lightcurve is about $0.92$
(modulo the uncertainty in lightcurve normalization, see Sec.~\ref{Sec:norm});
if the starspots were absolutely dark, they would still cover $8$~\% of the disk.
The largest peak-to-peak variation during one rotation spans from $0.98$ to $0.92$
indicating roughly $6$~\% more spots on the darker hemisphere.

Using a spot contrast of \mbox{$c_s = 0.7$},
which is about the average bolometric contrast of sunspots \citep{Beck1993, Chapman1994},
we determine a maximum spot coverage of $37$~\% on the eclipsed and $16$~\% on the noneclipsed section.
For the entire stellar surface a maximum and minimum spot coverage of $19$~\% and $16$~\% are derived.

\begin{figure}
    \includegraphics[height=0.5\textwidth,angle=-90]{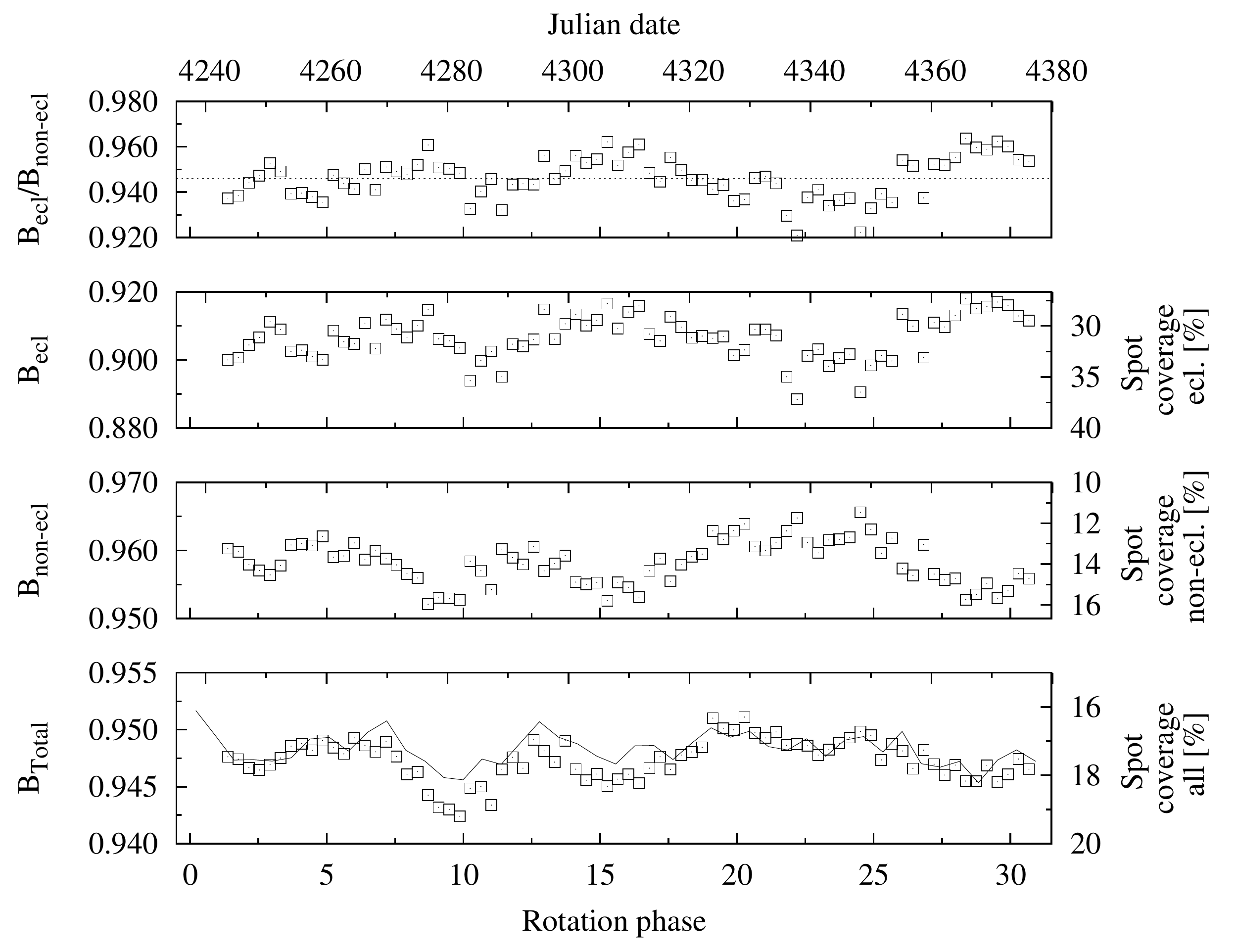}
    \caption{\label{Fig:brightnessMEAN}
             Average brightness values of the eclipsed ($B_{ecl.}$) and noneclipsed ($B_{non-ecl.}$) sections
             for each reconstruction interval.
             The y-axes on the right show the corresponding spot coverage for a spot contrast of $0.7$.
             \textit{Top panel}: Ratio $B_{ecl.}/B_{non-ecl.}$ of the mean brightnesses.
             \textit{Second panel from top}: Mean brightness and spot coverage for the eclipsed section.
             \textit{Third panel from top}: Mean brightness and spot coverage for the noneclipsed section.
             \textit{Bottom panel}: Total brightness \mbox{$B_{Total}$};
             the black curve shows the results of \citet{Lanza2009} shifted by $+9$~\%  (see Sect.~\ref{Sec:spot-cov}).}
\end{figure}

Details on brightness values and spot coverage can be found in Fig.~\ref{Fig:brightnessMEAN}.
The average brightness $B$ is the mean brightness of each reconstruction interval.
The top panel gives the ratio between the mean brightness of the eclipsed and noneclipsed sections,
the second and third panel (from top) the eclipsed and noneclipsed brightnesses separately,
and the bottom panel shows the variation of the total brightness
\mbox{$B_{Total} = (1-A) \cdot B_{ecl.} + A \cdot B_{non-ecl.}$} with \mbox{$A\,=\,0.79$}.
The associated spot coverage fraction is calculated using \mbox{$(1-B)/(1-c_s)$}.

Instead of converting brightness into spot coverage, we can also reverse the process.
Assuming the darkest element of the transit map is entirely covered by one spot,
the reconstructed brightness represents the spot contrast.
One strip on the eclipsed section has a size of about $1$~\% of
the entire surface.
We obtain a value of $0.76$ for the darkest surface element of the transit map,
which is not far from the solar spot contrast of $0.7$.
The minimum brightness of $0.76$ can be translated into a temperature contrast of \mbox{$~\sim 400$\textdegree~K} between
the spot and the photosphere, which is at \mbox{$T_p = 5625$\textdegree~K} \citep{Bouchy2008}.

\begin{figure}[t]
 \begin{center}
 \includegraphics[height=0.5\textwidth, angle=-90]{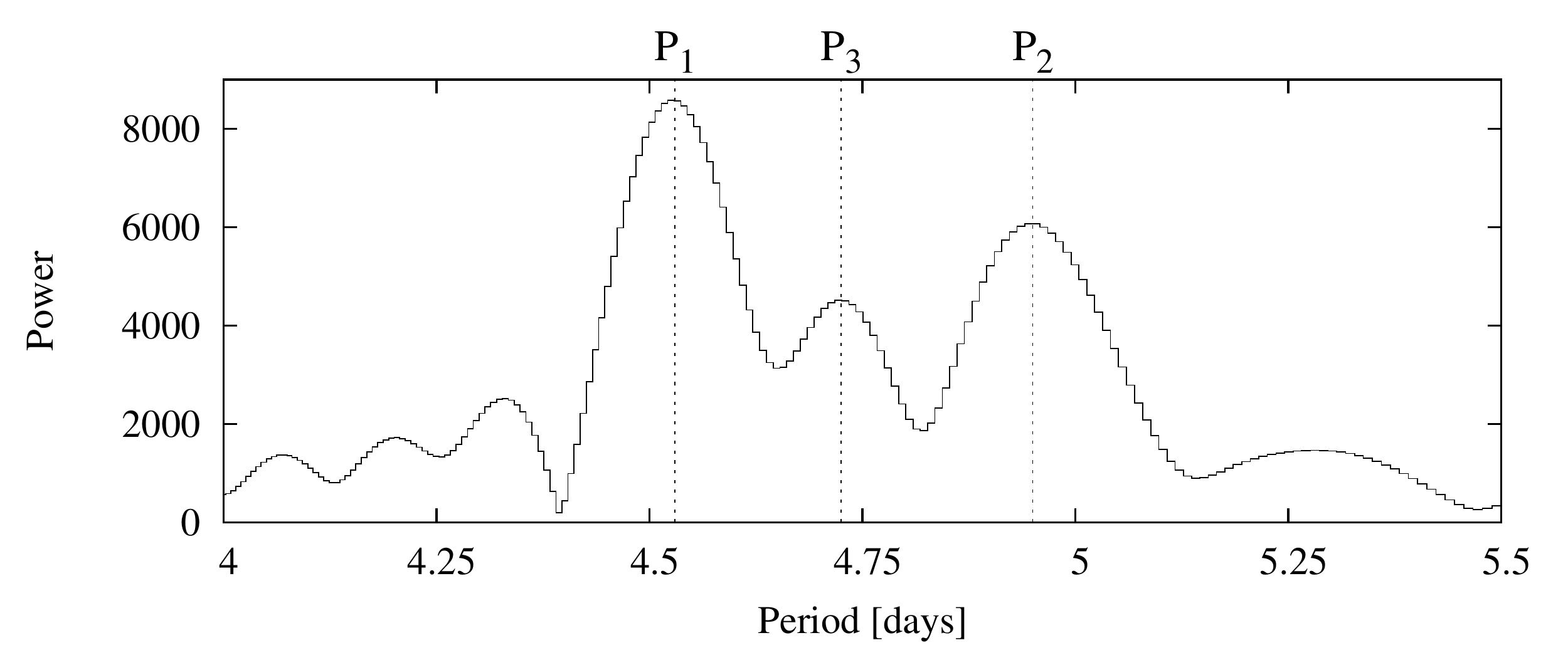}
 \end{center}
 \caption{\label{Fig:periodogram}
 Section of the Lomb-Scargle periodogram of the \mbox{\co-2} lightcurve containing the highest peak.
 The three peaks are labeled $P_1 = 4.53$, $P_2 = 4.95$, and $P_3 = 4.72$~days according to decreasing power.
 The adopted rotation period of our reconstructions, $P_s = 4.522$~days \citep{Lanza2009}, is close to $P_1$.}
\end{figure}

\subsection{Rotation period}

For \mbox{\co-2a} \citet{Lanza2009} determine a rotation period of \mbox{$(4.52\pm 0.14)$~d} by means of the Lomb-Scargle
periodogram, which the authors later refine to be \mbox{$P_s = (4.522\pm0.024)$~days} by minimizing the longitudinal migration
of their active longitudes.
We adopted $P_s$ for our reconstructions.
It is almost identical to the largest peak in the Lomb-Scargle periodogram (Fig.~\ref{Fig:periodogram})
centered at \mbox{$P_1 = 4.53$~d}.
$P_1$ is accompanied by two other distinguishable peaks at \mbox{$P_2 = 4.95$~d} and \mbox{$P_3 = 4.72$~d},
which are related to structures in our surface maps as discussed in Sect.~\ref{Sec:dis-dr}.

Since the planet crosses the stellar disk in a latitudinal band between $6$\textdegree \
and $26$\textdegree, spots on the eclipsed section must be located close to the equator.
The approximate vertical alignment of the darkest features in the transit map indicates that
low-latitude features rotate close to the adopted rotation period of $4.522$~days
used in our reconstructions.
In contrast, features on the noneclipsed section
do show longitudinal migration, which may be attributed to
differential rotation of spot groups at latitudes \mbox{$\apprge \, 30$\textdegree}.
This finding suggests that active regions close to the equator dominate the modulation of the lightcurve.

A closer inspection of the transit map 
shows a small but constant longitudinal shift of about $60$\textdegree \ over the entire $30$~rotations.
It is not only visible in the dark structures located at $\sim 200$\textdegree \
and $\sim 60$\textdegree \ longitude, but for the bright region at $\sim 300$\textdegree \ as well.
As a consequence, these low-latitude features do not rotate exactly with a rotation
period of $4.522$~days but about $2$\textdegree \ per rotation more slowly,
which translates into a rotation period of approximately $4.55$~days.
\citet{Lanza2009} state a retrograde migration of their second active longitude
corresponding to this rotation period.

We recalculated the surface reconstructions adopting a rotation period of $4.55$~days.
As expected the longitudinal shift previously detected in the transit map disappeared,
but some features now seem to migrate in the other direction.
Therefore, the exact rotation period of features on the eclipsed section is
probably slightly smaller than $4.55$~days.
The global map 
changes accordingly when using a rotation period of $4.55$~days;
the brightness distribution reconstructed at larger transit numbers is
shifted towards larger longitudes, but remains qualitatively the same, so that these maps
are not presented separately.

\begin{figure}[t]
 \begin{center}
 \includegraphics[scale=0.36]{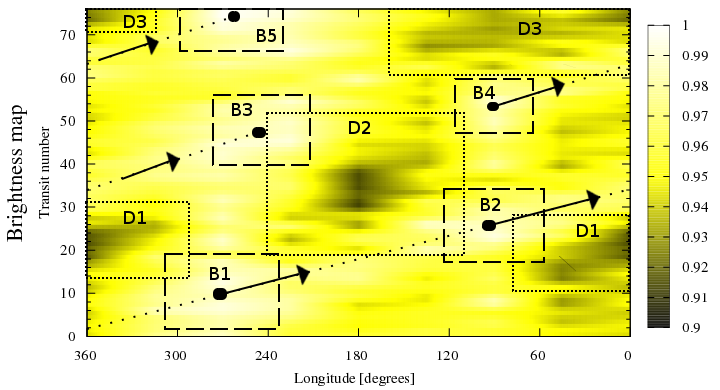}
 \end{center}
 \caption{\label{Fig:global-map-marked}
 The global surface map (left panel of Fig.~\ref{Fig:global-transit-maps})
 with pronounced dark and bright features marked and labeled \mbox{D1 -- D3} and \mbox{B1 -- B5}, respectively.
 The (green) dash-dotted line indicates the tentative movement of bright regions (see Sec.~\ref{Sec:dis-dr}).}
\end{figure}

\subsection{Longitudinal movement / differential rotation}
\label{Sec:dis-dr}

Some attributes of the global map 
suggest a longitudinal movement of surface structures;
a labeling of particularly interesting regions is given in Fig.~\ref{Fig:global-map-marked}.
Especially considering the inactive regions, the bright structure B1 at the bottom left
appears to move from $\sim 300$\textdegree \ to $\sim 100$\textdegree \ (B2) within
about $4$ stellar rotations. It continues moving in the direction of decreasing longitude
reaching roughly $240$\textdegree \ near transit number $50$ (B3).
Regions B4 and B5 do not seem to fit well into the line drawn by B1, B2, and B3.
However, a line connecting B4 and B5 roughly matches with the previously
detected rotation period between $4.8$ and $5.0$~days.
The tilted shape of the dark structure D1 roughly agrees with this range of periods as well.
D2 indicates a smaller rotation period consistent with $P_3$.

Although such surface map characteristics do not necessarily prove differential rotation,
they are certainly suggestive of differential rotation.
Assuming now this to be correct, we will elaborate on its consequences.
The largest rotation periods we obtain from structures of the global map lie between about $4.8$ and $5.0$~days.
We studied the longitudinal movements of these structures on surface maps obtained from lightcurve reconstructions
applying different rotation periods and always end up with similar results.
An examination of \co-2's periodogram reveals a splitting up of the highest
peak into three components:  \mbox{$P_1 = 4.53$~d}, \mbox{$P_2 = 4.95$~d},
and \mbox{$P_3 = 4.72$~d} (see Fig.~\ref{Fig:periodogram}).
It is striking that $P_2$ is close to the rotation period determined from tilted structures
in our global map.

If these rotation periods do arise from differential rotation,
we can estimate a lower limit of its strength.
The rotation period of eclipsed spots close to the equator is
consistent with the highest peak \mbox{$P_1 = 4.53$~days} of the periodogram,
the largest periods detected in the global map are around \mbox{$P_2 = 4.95$~days};
therefore, we adopt this period for the most slowly rotating active regions.
This way we determine a lower limit of \mbox{$\Delta \Omega > 0.1$ rad/d}
or \mbox{$\alpha > 0.08$}.
This is consistent with values expected for stars
with temperatures and rotation periods similar to that of \mbox{\co-2} \citep{Barnes2005}.
Using a 3-spot model approach,
\citet{Frohlich2009} derive an estimate of \mbox{$\Delta \Omega > 0.11$ rad/d},
which is in good agreement with our result.

Although the peaks in the periodogram fit nicely in with our brightness maps,
attributing them to three active regions with associated rotation periods may not be fully adequate.
We simulated several lightcurves with differentially rotating spots and examined their periodograms.
Although the main peak splits up into different components,
the periodogram does not necessarily map the exact rotation periods
of the differentially rotating spots to the peak barycenters.
An exhaustive analysis of the periodogram is beyond this work's scope, but
the above approach may serve as an approximation, and it shows that the characteristics of the
periodogram can be aligned with surface map attributes.

As an alternative, or maybe extension, to the interpretation in terms of differential rotation,
an evolution of the global activity pattern should be considered, which does not invoke longitudinal
movement of individual surface features, but a redistribution of strength between active regions.
The spot distribution on the global map suggests
sudden longitudinal relocations of the most active feature.
For the first $\sim 25$ transits the dominant spotted region D1
keeps its position at $\sim 30$\textdegree \ showing an apparent movement
towards smaller longitudes at the end of this interval.
Afterwards the dominant active region is found approximately $180$\textdegree \ apart from the previous position
in region D2.
Such $180$\textdegree-jumps, or rather `hemisphere-jumps', as the value of $180$\textdegree \ should not
be taken too seriously here, are found for inactive (bright) regions as well.

An appealing explanation for this apparent shift is a
change of the relative strength of two active regions, which does not involve movement of
any of the structures themselves.
These jumps are possibly a sign for some flip-flop scenario
as already claimed for other stars \citep{Jetsu1993, Korhonen2001},
where the relative strength between two active longitudes is changing suddenly and periodically.
The timescale of these `jump periods' derived from our global brightness map is roughly
$10$~stellar rotations; however, this is not seen in the transit map.

\begin{figure}[t]
 \begin{center}
 \includegraphics[height=0.46\textwidth, angle=-90]{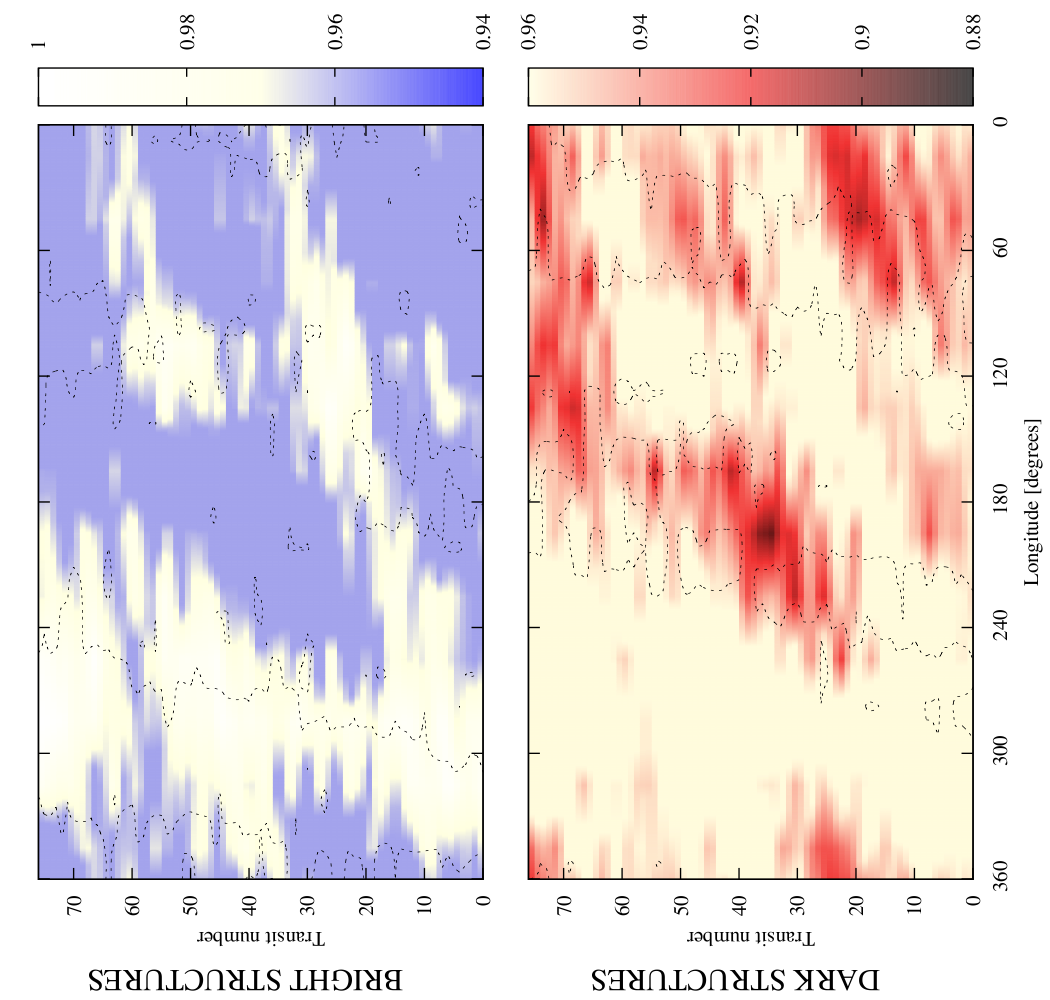}
 \end{center}
 \caption{\label{Fig:MAPcomp}
 Direct comparison of the two brightness maps of Fig.~\ref{Fig:global-transit-maps}.
 The global map is drawn in color, structures of the transit map are shown with contour lines.
 \textit{Upper panel:} Map of bright structures in the global map with \mbox{$b_{non-ecl.} > 0.95$}.
 The contour line delimits brightness values of the transit map above \mbox{$b_{ecl.} = 0.94$}.
 \textit{Lower panel:} Map of dark structures with \mbox{$b_{non-ecl.} < 0.95$}.
 Here the contour lines indicate a transit map brightness of \mbox{$b_{ecl.} = 0.87$}.
 }
\end{figure}

\subsection{Lifetimes of features}

Both the global and the transit map provide the possibility to measure
lifetimes of spotted regions.
For the noneclipsed section it is not clear whether a dark structure
actually is a single connected active region
or a superposition of several individual ones at roughly the same longitude
but completely different latitudes.

The transit map shows a high stability of the spot positions.
One group of spots is located between $200$\textdegree \ and $240$\textdegree,
which is stable up to transit number $35$;
then the spot coverage seems to decrease for about one stellar rotation.
Afterwards the spot distribution becomes more complicated and spreads over a larger area.
The left part of this more complex region, at about $210$\textdegree \ longitude,
is probably a continuation of the preceding active region;
however, at about $160$\textdegree~longitude a new, clearly separated group of spots appears.
This latter structure appears to have a lifetime of approximately $10$~stellar rotations,
which would be in good agreement with timescales observed on the global map.

Another group of spots is located between $0$\textdegree \ and $100$\textdegree.
At the beginning of the observation these spots are spread over this larger interval
of about $100$\textdegree \ on the eclipsed section;
later, approximately after transit number $\sim 50$, they seem to concentrate in a smaller interval.
This active region seems to have a significant fine structure in time,
which might indicate smaller lifetimes of individual spot groups;
however, those structures could also be artifacts of the reconstruction algorithm.
We prefer the interpretation that this active region,
and the left side of the other active region,
are persistent over all $30$~rotations,
although they probably contain smaller spots undergoing significant evolution on much shorter timescales.

The left side of both maps (around $300$\textdegree \ in longitude) shows bright regions which are least covered
by spots during all observations,
indicating a lifetime of half a year for this `inactive longitude'.
For the darkest structures of the global map, maximum lifetimes of about $10$ to $15$~stellar rotations
can be estimated. This is in agreement with the results of \citet{Lanza2009}, who determine a lifetime
of $\approx 55$~days ($=12$~stellar rotations) for active regions, and identify this time span
with the `beat period' visible in the lightcurve.

Although there are some structures in the transit map that suggest smaller lifetimes
than half a year for individual spot groups,
the lifetimes of features on the two different surface sections seem to be different.
A direct comparison of the global and the transit map is shown in Fig.~\ref{Fig:MAPcomp}.
The active regions on the eclipsed section remain active all the time
despite their possibly significant fine structure.
In contrast, active regions on the noneclipsed section evolve faster
showing more pronounced changes and a longitudinal movement
compared to the spots on the eclipsed section.
The reason for the apparent lifetime difference is not clear.
Possibly the darkest structures represent only a superposition of several spotted groups
at about the same latitude.
If these groups change their mutual longitudinal positions,
or a fraction of the spot groups dissolves,
the dark structures in our maps would brighten.
In this case the darkest structures would only represent special configurations
of the spot distribution and their `lifetimes' in our maps would not be
directly connected to the lifetimes of individual active regions on the surface.

The stable vertical alignment of features in our transit map cannot be caused
by a systematically incorrect transit profile used in our reconstructions.
The transits do not always cover exactly the same part of the eclipsed section,
which is also visible in the error distribution of the transit map;
dark structures indicate where the coverage is best, bright where it is worst.
Thus, an error introduced by the transit profile would be distributed
over the entire map.

\subsection{Comparing global and transit maps: spot migration?}

Finally, there is the possibility of detecting signatures of spot migration
-- the movement of spots from the equator to the poles or vice-versa --
in our brightness maps.
Figure~\ref{Fig:MAPcomp} presents the direct comparison of the brightest structures (upper panel)
and the darkest structures (lower panel) of the global and the transit map.
Especially the dark structure D2 (see Fig.~\ref{Fig:global-map-marked})
in the middle of the lower panel's map suggests
that features on the eclipsed and noneclipsed sections are related.
For the first $30$ transits there is a dark feature of the transit map at this longitude,
then it starts to disappear when D2 becomes darker.
This reflects a scenario where a spot group moves from the eclipsed section to
the noneclipsed.
After transit number $40$, D2 starts to disappear while other structures appear on the eclipsed section.
If this really represents a case of spot migration, the spot group either moves back onto the eclipsed section,
or it stays outside and new spots emerge under the transit path.


A similar observation can be made concerning the bright structures in the upper panel of Fig.~\ref{Fig:MAPcomp}.
The bright structures in the transit map between $60$\textdegree \ and $120$\textdegree \
alternate with the bright regions B2 and B4.
First there is a bright structure on the eclipsed map below region B2,
then there is a little bit of both between B2 and B4,
and after region B4 a bright structure is emerging in the transit map.

It is impossible to prove whether these signatures really represent spot migration;
probably some of them are due to other processes, e.g. short-term evolution of spotted regions.
Nevertheless, there is a similarity to what one would expect to see in brightness maps
from surfaces showing spot migration.
A behavior supporting a shift of spots from the eclipsed to the noneclipsed
sections (and vice-versa) can be observed in Fig.~\ref{Fig:brightnessMEAN} (second and third panel).
It suggests a correlation between the mean brightnesses of the two sections;
when the average brightness of the eclipsed section decreases, it increases on the noneclipsed part.
However, this correlation does not necessarily prove a steady motion between the two sections
and might as well indicate that vanishing spots just reappear somewhere else.

\begin{figure*}[t]
  \begin{center}
    \includegraphics[scale=0.35, angle=-90]{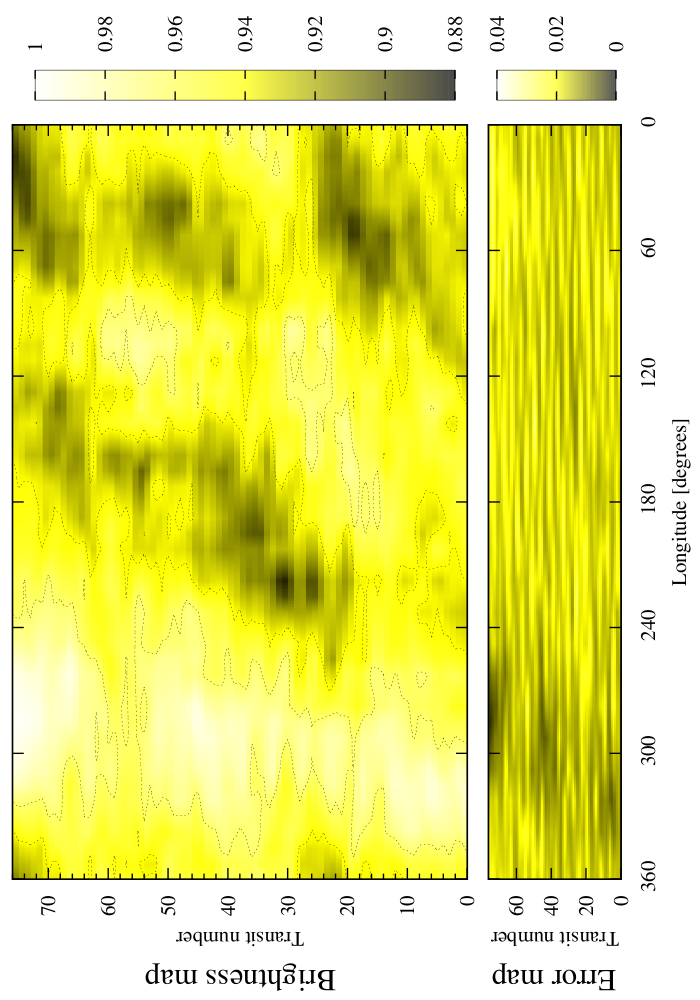}
    \includegraphics[bb= 60 60 620 980, clip, scale=0.27,angle=-90]{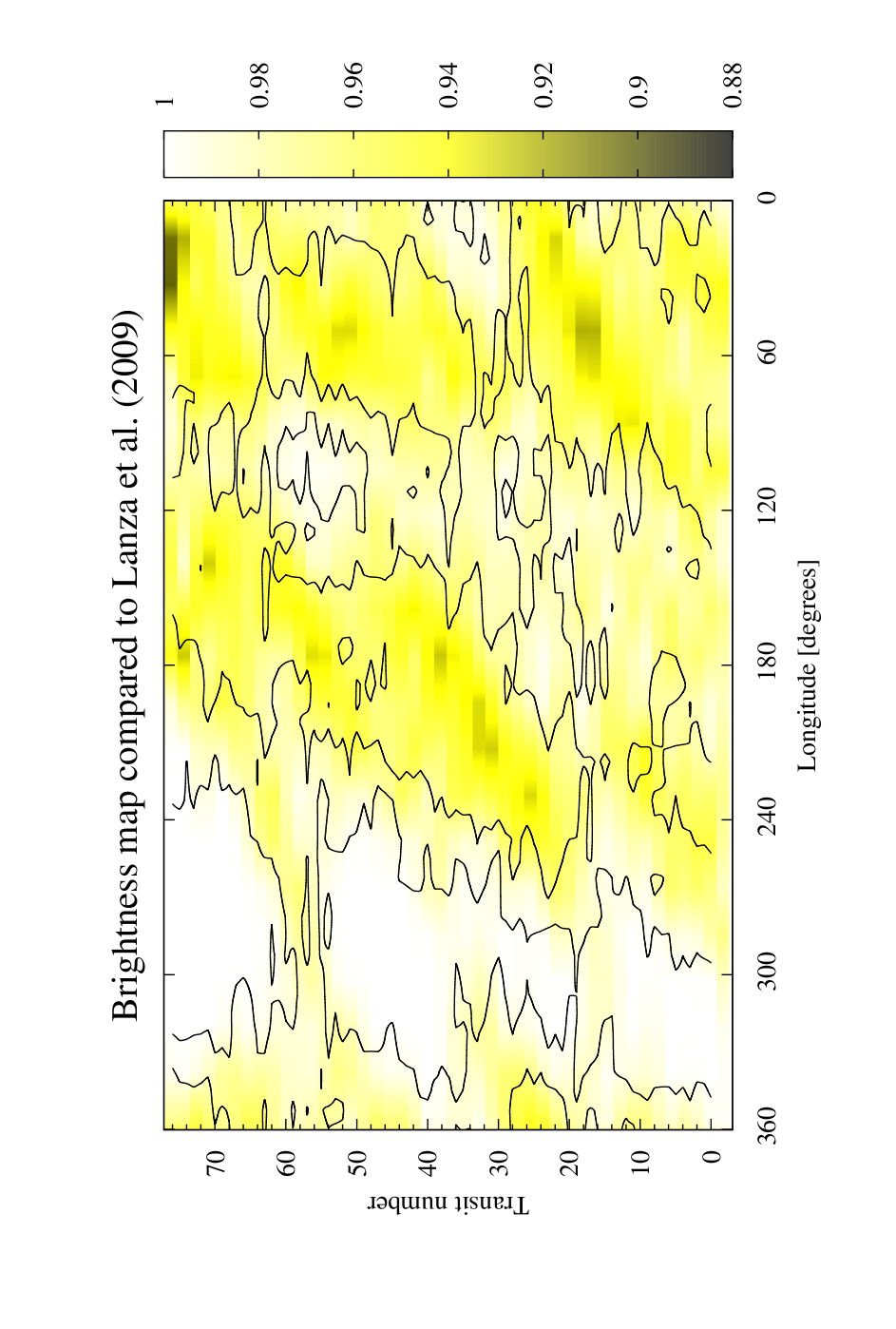}
  \end{center}
  \caption{\label{Fig:brightnessBOTH}
           \textit{Left panel}:
           Brightness distribution of the \textit{entire} surface for all reconstructed intervals;
           it represents a combination of both the global and the transit map.
           The error for each bin of the brightness map is shown below.
           \textit{Right panel}:
           Comparison of our brightness distribution (contours) to the reconstruction of \citet{Lanza2009},
           which are shown in color coding.
           See Sect.~\ref{Sec:comp-res} for details.
           }
\end{figure*}

\subsection{Comparison to previous results}
\label{Sec:comp-res}

Figure~\ref{Fig:brightnessBOTH} (left panel) displays the combined brightness map derived
from both the eclipsed (transit map) and noneclipsed sections (global map) of Fig.~\ref{Fig:global-transit-maps}:
the single maps are multiplied by their corresponding surface fractions
($0.21$ for the eclipsed and $0.79$ for the noneclipsed) and added.

\citet{Lanza2009} present a map of the surface evolution derived from a fit to the global lightcurve
(their Fig.~$4$) not including the transits.
In the right panel of Fig.~\ref{Fig:brightnessBOTH} we present a comparison of their results to ours.
Since we do not use filling factors, we translated their map into brightnesses
using their spot contrast of $0.665$.
We take the resulting map (color coding) and superimpose it on our combined map from both the eclipsed and
noneclipsed sections (contours).
In the left panel of Fig.~\ref{Fig:brightnessBOTH} the same contour lines are drawn
to provide a better comparison.
\citeauthor{Lanza2009}'s and our results show good agreement,
although a perfect match in fine-structure is neither found nor expected.
Dark and bright structures are located at very similar positions
and the shapes are consistent.

Adding up the brightness values of each reconstruction interval of the map in Fig.~\ref{Fig:brightnessBOTH} (left panel),
we can study the variations of the mean total brightness $B_{Total}$ of the star.
This is presented in Fig.~\ref{Fig:brightnessMEAN} (bottom panel).
With a maximum of $B_{Total} = 0.951$ and a minimum value of $0.942$,
the maximum difference between the highest and lowest average total brightness is only about $1$~\%,
which is much less than the maximum brightness differences within the brightness maps.
This implies that the star as a whole does not change its overall spot coverage as dramatically
as it redistributes it;
when spots disappear, other spots show up.
The solid line in the panel gives the comparison to the results of \citet{Lanza2009}.
We translated their values to our spot contrast of $0.7$ and shifted it
by a constant spot coverage of $+9$~\% to match our points.
Our average spot coverage of $17.5$~\% is roughly twice as high.
Although, a $9$~\% shift seems to be enormous, about $80$~\% of it ($+7$~\%) can be
attributed to a different normalization of the lightcurve and, thus, photosphere.
While \citet{Lanza2009} define the maximum flux in the lightcurve as their photospheric level,
our photosphere is $2$~\% brighter (cf. Sect.~\ref{Sec:norm}), which has to be compensated by spots.
We attribute the remaining $2$~\% to the differences in the adopted models.
In particular, we use longitudinal strips and spots can only be distributed
homogeneously across a strips, while \citet{Lanza2009} localize the spots in $200$~bins
on the surface.

Previously, we detected an average brightness under the eclipsed section
$(6\pm1)$~\% higher than on the noneclipsed section \citep{Huber2009}.
This value is redetermined from the reconstruction of the \textit{entire} lightcurve
presented in this paper.
It decreases to $(5.4\pm0.9)$~\%  (see top panel of Fig.\ref{Fig:brightnessMEAN}).

\subsection{Brightness maps and lightcurve modulation}


It is striking that the rotational variations of the star are
additionally modulated with a beat period about a factor $10$ to $15$ larger.
During the maxima of this large-scale modulation
(at about transit numbers $15$, $45$, and at the end of the lightcurve),
the minima of the stellar rotation are deep and regular,
i.e., the lightcurve has only one distinct minimum per rotation.
During the minima of this beat period (at about transit numbers $30$ and $60$),
the rotational modulation is flatter and more complex;
the minima are split up in two.
The beat period maxima indicate the existence of one large active region or longitude dominating the stellar surface.
The minima indicate that two smaller active regions at significantly
different longitudes imprint their signatures onto the lightcurve
leaving double-peaked structures during one stellar rotation.
This means dark regions are redistributed on timescales of $10$ to $15$~stellar rotations
from essentially one large feature to at least two smaller, longitudinally separated ones,
and then back to a large one.

This is also observable in our brightness maps.
Figure~\ref{Fig:brightnessBOTH} shows a change of the dominant surface feature
from $\sim 60$\textdegree \ to $\sim 220$\textdegree \ at about transit number~$20$.
Earlier the lightcurve was dominated by one large active longitude leading to
one broad minimum during each rotation.
During the transition phase, especially around transit number $25$,
the lightcurve minima become double-peaked.
After transit number $30$ the activity center of the active longitude at
$220$\textdegree \ moves to about $180$\textdegree \ and, thus, closer
to the other active longitude at about $60$\textdegree,
which has not entirely disappeared and starts to grow stronger again.
Because the active longitudes are closer now, the minima grow deeper;
additionally, the bright region at $300$\textdegree \ becomes larger
and the maxima rise.

Again there are double-peak structures between transit numbers $55$ and $60$
for the same reasons.
The decrease of the strong maxima is primarily due to the temporary change of the
bright longitude at $300$\textdegree.
Interestingly, the amplitude and width of the minima become even larger
at the end of the lightcurve, where the active longitudes first move closer together.
In the end, the feature at $\sim 20$\textdegree \ becomes dominant
spreading over almost $60$\textdegree \ in longitude and leading to very low lightcurve minima.

The main reason why the rotation period of the star is so nicely indicated by
almost constantly separated maxima of the lightcurve --
despite the pronounced surface evolution --
is that the on average brightest part of the surface,
located on the inactive longitude at $300$\textdegree,
remains stable at its position.

\section{Conclusions}

We present a reconstruction of the complete \co-2 lightcurve -- including transits --
covering about $140$~days or $30$~stellar rotations.
In contrast to previous work,
both the transit profiles and the rotationally-modulated global lightcurve are fitted simultaneously
leading to a \textit{consistent} solution for the entire lightcurve.
From the transits the brightness distribution on the eclipsed surface section is recovered,
which is the part of the surface between $6$\textdegree \ and $26$\textdegree \ latitude
constantly eclipsed by the planet.
The noneclipsed section is reconstructed from the rotational modulation.

The evolution of the spot distributions on both surface sections
is presented in two maps showing the surface brightness distribution as a function of time.
The composite map, which shows the evolution of the entire surface, is juxtapositioned with
previously published results. The results are found to be in agreement taking into account
different modeling approaches and assumptions.

The transit map shows two preferred longitudes densely covered with spots and
separated by approximately $180$\textdegree.
Both active regions persist for the entire observing time of $\sim 140$~days,
although they undergo significant evolution in size, structure, and brightness.
In the transit map they show a constant retrograde movement indicating
that the adopted stellar rotation period of \mbox{$P_s = 4.522$~days}
does not exactly describe their rotation.
We determine that these low-latitude features rotate
at a period of approximately $4.55$~days, which is
also true for the long-lived inactive longitude at $\sim 300$\textdegree.

The global map is more complex than the transit map
presumably because the structures it describes represent a superposition of features
at similar longitudes but different latitudes.
Usually it shows only one dominant dark feature at a time,
which changes position after approximately $10$ to $15$~stellar rotations.
Again these dominant features are separated by about $180$\textdegree \ in longitude.
A persistent inactive longitude exists at about $300$\textdegree \
similar to the one in the transit map and at about the same position.
The global map indicates that there are features located on the noneclipsed section
with significantly larger rotation periods than $4.522$~days.
This suggests the presence of differential rotation
with spots moving more slowly at high-latitudes than at low-latitudes.
We estimate a differential rotation of $\Delta \Omega > 0.1$ or $\alpha > 0.08$, respectively.

Assuming a spot contrast of $0.7$,
the spot coverage of the eclipsed section reaches a maximum of $37$~\%,
which is more than twice as large as the maximum on the noneclipsed section.
\textit{On average} the eclipsed section is ($5 \pm 1$)~\% darker
than its noneclipsed counterpart.
Sunspots are located within $\pm 30$\textdegree \ around the solar equator.
Similarly, our results indicate that spot groups on \co-2 are also concentrated in a low-latitude `active belt'.

\begin{acknowledgements}
We thank A.~F.~Lanza for kindly providing us with their results for comparison.
K.H. is a member of the DFG Graduiertenkolleg 1351 \textit{Extrasolar Planets and their Host Stars}
and acknowledges its support.
S.C. and U.W. acknowledge DLR support (50OR0105).
\end{acknowledgements}


\bibliographystyle{aa}
\bibliography{referenz}


\end{document}